\documentclass[12pt,french]{article}

\usepackage{latexsym,amsmath,amssymb,fancybox}

\usepackage[french]{babel}

\usepackage{color,epsfig}
\usepackage{graphicx}

\definecolor{navy}{rgb}{0.0,0.0,0.4}

\definecolor{rd}{rgb}{1,0,0}
\definecolor{or}{rgb}{1,.33,0}
\definecolor{pi}{rgb}{.66,.33,.33}
\definecolor{gn}{rgb}{0,.50,0}
\definecolor{be}{rgb}{0,0,.66}
\definecolor{ma}{rgb}{.66,0,.33}
\definecolor{vi}{rgb}{.33,0,.66}
\definecolor{gy}{rgb}{0,.33,.66}
\definecolor{ye}{rgb}{.66,.33,0}
\definecolor{bk}{rgb}{0,0,0}

\font\sm=cmr9

\font\sixrm=cmr6

\font\impa = cmssbx10 scaled \magstep3

 \font\fmi=cmitt10 scaled \magstep2

\def\thf{\baselineskip=\normalbaselineskip\multiply\baselineskip
by 7\divide\baselineskip by 6}

\def\fff{\baselineskip=\normalbaselineskip}

\def\spose#1{\hbox to 0pt{#1\hss}}
\def\lta{\mathrel{\spose{\lower 3pt\hbox
{$\mathchar"218$}}\raise 2.0pt\hbox{$\mathchar"13C$}}}  \def\gta{\mathrel
{\spose{\lower 3pt\hbox{$\mathchar"218$}}\raise 2.0pt\hbox{$\mathchar"13E$}}}

\def\Libra{\spose {--} {\cal L}}
\def\Euro{\spose {\lower 2.5pt\hbox{${^{\bf =}}$}}{ C}}

\def\spose#1{\hbox to 0pt{#1\hss}}

\def\sqr#1#2{{\vcenter{\hrule height.4pt\hbox{\vrule width.8pt height#2pt
\kern#1pt\vrule width.8pt}\hrule height.4pt}}}

\begin{document}

\def\bb{\large\color{bk}  $ }
\def\fb{ $  }
\def\be{\large\begin{equation}\color{bk} }
\def\fe{\end{equation}}

\newcommand{\eqn}{\label}
\newcommand{\bel}{\begin{equation}\label}

\def\eqdef{\fff\ \vbox{\hbox{$_{_{\rm def}}$} \hbox{$=$} }\ \thf }

\def\ov{\overline}


\def\eu{{\color{bk}\,\hbox{\fmi e}\,}}

\def\d{\delta}\                   \def\dr{{\rm d}}
\def\dL{\spose {\lower 5.0pt\hbox{\sixrm \color{gn}L} } {\delta}}
\def\dE{\spose {\lower 5.0pt\hbox{\sixrm \color{be}E} } {\delta}}
\def\dP{\spose {\lower 2.0pt\hbox{$_{_\Gammab}$} } {\delta}}
\def\deI{\spose {\raise 3.0pt\hbox{$\, \acute{\ }$}} {\delta}}
\def\deJ{\spose {\raise 3.0pt\hbox{$\, \grave{\ }$}} {\delta}}

\def\alphar{{\color{bk} \alpha}}
\def\chir{{\color{bk} \chi}}


\def\nb{{\rm\color{be} n}}
\def\pb{{\rm\color{be} p}}
\def\db{{\rm\color{be} d}}
\def\ud{{^{\,_{(\db)}}}\!}  
\def\udi{{^{\,_{( \db-1)}}}\!}
\def\up{{^{\,_{( \pb)}}}\!}  

\def\gb{{\color{be} g }} 
\def\nabl{{\color{be}\nabla\!}}
\def\nablab{ {\color{be} \nabla}}
\def\bepsilon{{\color{be}\varepsilon}}
\def\calR{{\color{be}{\cal R}}} 
\def\calW{{\color{be}{\cal W}}}
\def\calS{{\color{be}{\cal S}}}
\def\Gammab{ {\color{be} \Gamma}}

\def\phib{{\color{be} \phi}}
\def\hb{{\color{be} h}}
\def\nab{{\color{be}\nabla}}
\def\xib{{\color{be} \xi}}


\def\gamr{{\color{ma}\gamma}}
\def\velr{{\color{ma}v }}
\def\thetr{{\color{ma}\theta}} 
 \def\Lhat{\hat{\color{ma}L}}

\def\Fr{{\color{ma} F}}
\def\Nr{{\color{ma} N}}   

\def\xm{{\color{ma} x}}
\def\omegam{{\color{ma} \omega}}

\def\rmA{ {_{\rm \color{ma}A}}} 
\def\alphab{ {\color{ma}\alpha} }

\def\Thetar{ {\color{ma}\Theta} } 
\def\varthetar{ {\color{ma}\vartheta} }

\def\Omegar{ {\color{ma}\Omega} } 
\def\varpir{ {\color{ma}\varpi} }
\def\varpirIJ{\spose {\raise 3.0pt\hbox{$\, \acute{\ }\grave{\ }$}}{\varpir}}

\def\OP{ {\color{ma}{\mathfrak O}}}

\def\Lhatr{\hat{\color{ma}L}}
\def\Lambdahatr{\hat{\color{ma}\Lambda}}
\def\jhatr{\hat{\color{ma}j}}
\def\Whatr{\hat{\color{ma}{\,\hbox{\fmi w}\,}}}
\def\Thatr{\hat{\color{ma}T}}


\def\Av{{\color{vi} A}} 
\def\Bv{{\color{vi} B}}


\def\varphiv{{\color{vi}\varphi}} 
\def\qv{ {\color{vi} q} }

\def\uvec{ {\color{gn} u}}
\def\uvert{ {\color{vi} u}}
\def\vww{ {\color{vi} w} }
\def\vv{ {\color{vi} v} }
\def\bv{ {\color{vi} b} }
\def\Ggamma{ {\color{vi} {\mit\Gamma}} }
\def\xxxi{ {\color{vi} \xi} }
\def\Ddelta{ {\color{vi} {\mit\Delta}} }


\def\lambdag{{\color{gn}\lambda}} 
\def\perpg{{\color{gn}\perp\!}}    
\def\etag{{\color{gn}\eta}}
\def\hg{{\color{gn}\eta}}
\def\gammagn{{\color{gn}\gamma}}

\def\onab{{\color{gn}\ov\nabla\!}}
\def\nablag{ {\color{gn} \nabla}}
\def\Kg{{\color{gn} K}} 
\def\Re{{\color{gn} R}}

\def\Cg{{\color{gn}\,\hbox{\fmi C}\,}}
\def\Omegag{{\color{gn}\mit\Omega}} 
\def\Xe{{\color{gn}\mit\Xi}} 
\def\calEg{{\color{gn}\cal E}} 
\def\calSg{\ov{\color{gn}\cal S}}
\def\calC{{\color{gn}\cal C}}    

\def\ugn{{\color{gn}\,\hbox{\fmi u}\,}} 
\def\vgn{{\color{gn}\,\hbox{\fmi v}\,}}

\def\varepsilong{ {\color{gn} \varepsilon}}
\def\ebl{ {\color{gn} e}}
\def\nN{ {\color{gn} n}}
\def\vN{ {\color{gn} N} }
\def\aa{ {\color{gn} a} }

\def\xig{{\color{gn}\xi}}
\def\ug{{\color{gn} u}}
\def\ag{{\color{gn} a}}
\def\Delg{{\color{gn} \Delta}}

\def\ii{{\color{gy} { i}}}    
\def\ji{{\color{gy} { j}}} 
\def\nauti{{\color{gy}_0}} \def\wuni{{\color{gy}_1}} 
\def\prim{{\color{gy}\prime}} 
\def\dit{{\color{gy}\dot{\,}\!\!}} \def\ddit{{\color{gy}\ddot{\,}\!\!}}
\def\sigme{{\color{gy}\sigma}} 
\def\Kie{{\color{gy}\,\hbox{\fmi K}\,}}
\def\rhoe{{\color{gy}\rho}} \def\pome{{\color{gy}\varpi}}
\def\lambde{{\color{gy}\lambda}} \def\iote{{\color{gy}\iota}} 
\def\A{{\color{gy}_A}} \def\B{{\color{gy}_B}}
\def\X{{\color{gy}_X}} \def\Y{{\color{gy}_Y}}


\def\calL{ {\color{rd}{\cal L}} }    
\def\calH{ {\color{rd}{\cal H}} }
\def\calT{ {\color{rd}{\cal T}} }
\def\calU{ {\color{rd}{\cal U}} }

\def\Lr{ {\color{rd} {L}} }
\def\olag{{\color{rd}\ov{L}}} 
\def\duLag{ {\color{rd} {\Lambda}} }

\def\of{{\color{rd}\ov{f}}} 
\def\tf{{\color{rd}f}}
\def\fex{ {\color{rd} f} } 
\def\cf{{\color{rd}\check{f}}}

\def\oj{{\color{rd}\ov {\,j}}}  
\def\oW{{\color{rd}\ov{\,\hbox{\fmi w}\,}}}
\def\bet{{\color{rd}\beta}}  

\def\Tr{{\color{rd} T }} 
\def\Ur{{\color{rd} U }} 
\def\oT{{\color{rd}\ov{T}}}

\def\cE{{\color{rd}\hbox{\fmi c}_{_{\rm E}}}} 

\def\mur{{\color{rd}\mu}} \def\nur{{\color{rd}\nu}}
  
\def\ze{{\color{rd}\,\hbox{\fmi z}\,}}

\def\gamE{{\color{rd}\gamma_{_{\rm E}}}}

\def\mur{{\color{rd}\mu}} 
\def\nue{{\color{rd}\nu}}
\def\sen{{\color{rd} s}} 
\def\Thete{{\color{rd}{\mit\Theta}}}
\def\hen{{\color{rd} h}}

\def\rkap{ {\color{rd} \kappa} }

\def\ocr{ \overline{\color{rd} c} }
\def\Cr{ {\color{rd} C} }
\def\Jr{ {\color{rd} J} }
\def\PPi{ {\color{rd} {\mit \Pi}} }
\def\Upsi{ {\color{rd} {\mit \Upsilon}} }

\def\pred{ {\color{rd}p} }          
\def\pired{ {\color{rd}\pi} }
\def\Cred{ {\color{rd} {\mathfrak C} } }           

\def\omegar{{\color{rd} \omega}}
\def\jr{{\color{rd}j}}
\def\Wr{{\color{rd}{\,\hbox{\fmi w}\,}}}
\def\Lr{{\color{rd} L}}
\def\Lambdar{{\color{rd} \Lambda}}
\def\Cauch{ {\color{rd} {\mathfrak C} } }
\def\kappar{{\color{rd} \kappa}}


\def\ee{{\color{or}\hbox{\fmi e}}}  \def\qe{{\color{or}\hbox{\fmi q}}}  
\def\mag{{\color{or}\,\hbox{\fmi m}\,}}  
\def\Ze{{\color{or}\,\hbox{\fmi Z}\,}} 
\def\Ne{{\color{or}\,\hbox{\fmi N}\,}}
\def\Action{{\color{or}\cal I}}

\def\mm{ {\color{or} m} }
\def\delth{ {\color{or} \delta} }
\def\MM{ {\color{or} {M}} }
\def\bern{ {\color{or} b} }

\def\smM{{\color{or}\sm M}}   
\def\Msr{{\hbox{\smM}}}
\def\smG{{\color{or}\sm G}}
\def\smQ{{\color{or}\sm Q}}   
\def\Qsr{{\hbox{\smQ}}}
\def\ffr{{\color{or}\,\hbox{\fmi f}\,}}

\sf 
\color{navy}

\begin{center}
 {\textcolor{ma}
 {\impa Classical dynamics of strings and branes,
  \\ with application to vortons}}    \\[0.8cm]
 \underline{Brandon Carter} \\[0.4cm]
 \textcolor{gn}{LUTH (CNRS), Observatoire Paris - Meudon }
  \\[0.4cm]
  {\color{be}{\it Invited contribution. to Proc. {\bf JGRG20}, \\
ed. T. Hiramatsu, M. Sasaki, M. Shibata, T, Shiromizu, 
\\ Yukawa Institute, Kyoto, September, 2010.}}\\[0.8cm]
\end{center}

 {\bf Abstract}. These notes offer an introductory overview of
the essentials of classical brane dynamics in a space-time 
background of arbitrary dimension, using a systematic geometric 
treatment emphasising the role of the second fundamental tensor
and its trace, the curvature vector {\bb \Kg^\mu\fb}.
This approach is applied to the problem of stability of vorton 
equilibrium states of cosmic string loops in an ordinary 4-dimensional
background.

\bigskip


 \section{Worldsheet Curvature Analysis}\label{SectI}


\subsection{The first fundamental tensor} \label{1-1}

Earlier treatments of the classical dynamics of strings
and higher p-branes were inclined to rely too much on gauge dependent
auxiliary structures such as internal coordinates {\bb \sigme^\ii\fb}
 on the \db=p+1 dimensional worldsheet, which can be useful
for various computational purposes but tend to obscure what is essential.
The present notes offer an introductory overview of a more geometrically
elegant approach \cite{C00} that is particularly useful for work
in a background spacetime whose dimension \nb\, is 5 or more 
\cite{Niarchos08,Cao09,Emparan10}, but that I originally 
developed for the purpose of studying
cosmic string loops and particularly the question of the stability
of their vorton equilibrium states \cite{CarMar93}
in a background of dimension \nb=4. Following the strategy
originally advocated by Stachel\cite{Stachel80}, the guiding 
principle of this approach \cite{C00} is to work
as far as possible with a
single kind of tensor index, which must of course be the one that is most
fundamental, namely that of the \nb -dimensional 
coordinates, {\bb x^\mu\, ,\fb} on the background spacetime with
metric {\bb \gb_{\!\mu\nu}\, .\fb} 

The idea is to avoid unnecessary use of the internal coordinate 
indices, which are lowered and raised by contraction
with the induced metric 
{\bb \hg_{\ii\ji}=\gb_{\!\mu\nu} x^\mu_{\ ,\ii} x^\nu_{\ ,\ji}
\fb} 
(using the notation
{\bb x^\mu_{\ ,\ii}={\partial x^\mu}/{\partial\sigme^\ii}\fb})
on the worldsheet, and with its contravariant inverse 
{\bb \hg^{\ii\ji}\, .\fb}
This is achieved by working instead with the (first) fundamental 
tensor as given by projection back onto the background according 
to the prescription
{\be \etag^{\mu\nu}= \hg^{\ii\ji} x^\mu_{\ ,\ii} x^\nu_{\ ,\ji} \, ,
\eqn{1.5}\fe}
(in the manner that is  applicable to 
the contravariant version of any worldsheet tensor) so that
{\bb \etag^\mu_{\ \nu}\fb} will be the tangential projector.
The complementary orthogonal projector is
{\bb \perpg^{\!\mu}_{\,\nu}=\gb^\mu_{\ \nu}-\etag^\mu_{\ \nu}\, .
\fb}
As well as having the  properties 
{\bb \etag^\mu_{\ \rho}\,\etag^\rho_{\ \nu}=\etag^\mu_{\ \nu} \, ,\fb}
and
{\bb\perpg^{\!\mu}_{\ \rho}\perpg^{\!\rho}_{\ \nu}=\perpg^{\!\mu}_{\ \nu}
\fb} 
these projection tensors 
will evidently be related
by 
{\bb \etag^\mu_{\ \rho}\perpg^{\!\rho}_{\ \nu}\,=\,0\,=\perpg^{\!\mu}_{\ \rho}
\etag^\rho_{\ \nu} \, . 
\fb}





\subsection{The second fundamental tensor }
\label{1-3}

In so far as we are concerned with tensor fields such as the frame vectors
whose support is confined to the \db-dimensional world sheet, the effect of
Riemannian covariant differentation {\bb \nabl_\mu\fb} along an arbitrary
directions on the background spacetime  will not be well defined, only the
corresponding tangentially projected differentiation operation
{\be \onab_\mu\eqdef\etag\,^\nu_{\ \mu}\nabl_\nu\, , \eqn{1.7}\fe}
being meaningful for them, as for instance in the case of a scalar field
{\bb \varphiv\fb} for which the tangentially projected gradient is given in
terms of internal coordinate differentiation simply by
{\bb \onab{^\mu}\varphiv=\hg^{\ii\ji} x^\mu{_{\!,\ii}}\,\varphiv_{,\ji}\, .\fb}
The action of this operator on
the first fundamental tensor {\bb \etag^{\mu\nu}\fb} itself gives
 the entity
{\be \Kg_{\mu\nu}{^\rho} \eqdef \etag\,^\sigma_{\ \nu}\onab_\mu
\etag\,^\rho_{\ \sigma} \eqn{1.10}\fe}
that we refer to \cite{C00} as the {\it  
second fundamental tensor}.


As this second fundamental tensor, {\bb \Kg_{\mu\nu}{^\rho}\fb} will play an 
important role in the work that follows, it is worth lingering 
\cite{C00} over its 
essential properties. The expression (\ref{1.10}) could of course be 
meaningfully applied not only to the fundamental projection tensor of a 
d-surface, but also to any (smooth) field of rank-d projection operators 
{\bb \etag\,^\mu_{\ \nu}\fb} 
as specified by a field of arbitrarily orientated d-surface elements. What
distinguishes the integrable case --  in which the elements mesh
together to form a well defined d-surface through the point under
consideration -- is the {\it Weingarten identity}, whereby 
that the tensor defined by (\ref{1.10}) will have the symmetry property
{\be \Kg_{[\mu\nu]}{^\rho} =0 \, ,\eqn{1.13}\fe} 
an integrability condition that is  derivable \cite{C00} 
as a version of the well known Frobenius theorem. 



As well as being symmetric, the  tensor  {\bb \Kg_{\mu\nu}{^\rho}\fb} is  
obviously tangential on the first two indices and
also orthogonal on the last: 
{\bb\perpg^{\!\sigma}_{\,\mu}\Kg_{\sigma\nu}{^\rho}=\Kg_{\mu\nu}{^\sigma}
\etag_\sigma{^\rho}=0   \, . 
\fb}
It  fully determines
 the tangential derivatives of the first fundamental tensor
{\bb \etag\,^\mu_{\ \nu}\fb} by the formula
{\be \onab_\mu\etag{_{\nu\rho}}=2\Kg_{\mu(\nu\rho)}\, , \eqn{1.15}\fe}
(using round brackets to denote symmetrisation) and it is
characterisable by the condition that the orthogonal projection of the
acceleration {\bb \dot \ugn^\mu={\ugn^\nu\nabl{_\nu} \ugn^\mu} \fb}
of any tangential unit vector field {\bb\ugn^\mu\fb} 
(with {\bb \ugn^\mu \ugn_\mu=-1\fb}) will be given by
{\bb \ugn^\mu \ugn^\nu \Kg_{\mu\nu}{^\rho}
=\perpg^{\!\rho}_{\,\mu}\dot \ugn^\mu\, .
\fb}


\subsection{Extrinsic curvature vector and Conformation 
tensor}
\label{1-4}

It is very practical for a great many purposes to introduce the {\it
extrinsic curvature vector} {\bb \Kg^\mu\, ,\fb} defined 
\cite{C00} as the trace of the second fundamental tensor, 
{\be \Kg^\mu\eqdef \Kg^\nu_{\ \nu}{^\mu}=\onab_\nu\etag^{\mu\nu}
\ , \eqn{1.17}\fe}
which is automatically orthogonal to the worldsheet, 
 {\bb\etag^\mu_{\ \nu}\Kg {^\nu}=0 \, .\eqn{1.17a}\fb}
It is useful for many specific purposes to work this out in terms of the
intrinsic metric {\bb \hg_{\ii\ji}\fb}  and its determinant 
{\bb \vert\hg\vert \fb}.  
For the tangentially projected gradient of a scalar field $\varphiv$ on the
worldsheet, it suffices to use the simple expression 
{\bb\onab^{\,\mu}\varphiv=\hg^{\ii\ji} x^\mu{_{,\ii}}
\varphiv_{,\ji}\, .\fb}
However for a tensorial field (unless one is using Minkowski coordinates 
in a flat spacetime) the gradient will also have contributions involving
the background Riemann Christoffel connection 
{\bb\Gammab_{\mu\ \rho}^{\,\ \nu}=\gb^{\nu\sigma}\big(\gb_{\sigma(\mu,\rho)}
-\frac{_1}{^2}\gb_{\mu\rho,\sigma}\big)\, .
\fb}
The curvature vector is thus obtained in explicit detail as
{\be \Kg^\nu={1\over\sqrt{\Vert \hg\Vert}}\Big(
\sqrt{\Vert\hg\Vert}\hg^{\ii\ji}x^\nu_{\, ,\ii}\Big){_{,\ji}}+
\hg^{\ii\ji}x^\mu_{\, ,\ii}x^\rho_{\, ,\ji}\Gamma_{\mu\ \rho}^{\,\ \nu}\, .
\eqn{1.19}\fe}
This expression is useful for specific computational purposes, but much of 
the literature on cosmic string dynamics has been made unnecessarily heavy 
by a tradition of working all the time with long strings of non tensorial 
terms such as those on the right of (\ref{1.19}) rather than exploiting more 
succinct tensorial expressions, such as \  
{\bb \Kg^\nu=\onab_\mu\etag^{\mu\nu}\, .\fb}


 As an alternative to the universally applicable tensorial approach advocated 
here, there is of course another more commonly used method of achieving 
succinctness in particular circumstances, which is to sacrifice gauge 
covariance by using specialised kinds of coordinate system. 
In particular, for the case of a string, i.e. for a 2-dimensional 
worldsheet, it is standard practise to use conformal coordinates 
{\bb \sigme^{\nauti}\fb}  and {\bb \sigme^{\wuni}\fb} so that the 
corresponding tangent vectors {\bb\, \dit x^\mu=x^\mu_{\, ,\nauti}\fb} and 
{\bb x^{\prim\mu}= x^\mu_{\, ,\wuni}\fb}  satisfy the restrictions 
{\bb\, \dit x^\mu x^\prim_{\, \mu}=0\, ,\fb} 
{\bb \dit x^\mu\dit x_\mu+x^{\prim\mu}x^\prim_{\,\mu}=0\, ,\fb} which implies 
{\bb \sqrt{\Vert \hg\Vert}=x^{\prim\mu}x^\prim_{\,\mu}=
-\dit x^\mu\dit x_\mu\, ,\fb} so that (\ref{1.19}) simply gives 
{\bb  \sqrt{\Vert\hg \Vert}\,\Kg^\nu=\fb}
{\bb x^{\prim\prim\nu}-\ddit x^\nu + (x^{\prim\mu}x^{\prim\rho}
-\dit x^\mu\dit x^\rho)\Gamma_{\mu\ \rho}^{\,\ \nu}\, .\fb}


The physical specification of the extrinsic curvature vector (\ref{1.17}) 
for a timelike \db-surface in a dynamic theory provides what can be taken 
as the equations of extrinsic motion of the d-surface 
\cite{C00}, the simplest possibility being 
the ``harmonic'' condition $\Kg^\mu=0$ that is obtained (as shown below) 
from a surface measure variational principle such as that of the Dirac 
membrane model \cite{Dirac62}, or of the Goto-Nambu string model 
\cite{Kibble76} whose dynamic equations in a flat
background are therefore expressible with respect to a standard conformal
gauge in the familiar form {\bb x^{\prim\prim\mu}-\ddit x^\mu=0\, ,\fb} 

There is a certain analogy between the Einstein vacuum equations, which
impose the vanishing of the trace {\bb \calR_{\mu\nu}\fb} of the background
spacetime curvature {\bb \calR_{\lambda\mu}{^\rho}{_\nu}\, ,\fb} and the
Dirac-Gotu-Nambu equations, which impose the vanishing of the trace
{\bb \Kg^\nu\fb} of the second fundamental tensor 
{\bb \Kg_{\lambda\mu}{^\nu}\, ,\fb}
Moreover, just as it is useful to separate out the Weyl tensor
\cite{Schouten54}, i.e. the trace free part of the Ricci background
curvature which is the only part that remains when the Einstein vacuum
equations are satisfied, so also analogously, it is useful to separate
out the the trace free part of the second fundamental tensor, namely
the extrinsic conformation tensor \cite{C00}, which is the only
part that remains when equations of motion of the Dirac - Goto - Nambu
type are satisfied. 

Explicitly, the trace free {\it extrinsic
conformation} tensor {\bb \Cg_{\mu\nu}{^\rho}\fb} of a \db-dimensional 
imbedding is defined \cite{C00} in terms of its first 
and second fundamental tensors 
as
{\be \Cg_{\mu\nu}{^\rho}\eqdef \Kg_{\mu\nu}{^\rho}-{1\over{\db}}
\etag{_{\mu\nu}} \Kg^\rho \ , \hskip 1 cm     
\Cg^\nu_{\ \nu}{^\mu}=0 \ .\eqn{1.20}\fe}



Like the Weyl tensor $\calW_{\lambda\mu}{^\rho}{_\nu}$ of the background 
metric (whose  definition is given implicitly by (\ref{1.25}) below) this
conformation tensor has the noteworthy property of being invariant with
respect to conformal modifications of the background metric:
{\bb \gb_{\mu\nu}\mapsto {\rm e}^{2\alpha}\gb_{\mu\nu}\ 
 \hskip 0.6 cm
\Rightarrow\fb}
{\be \Kg_{\mu\nu}{^\rho}\mapsto \Kg_{\mu\nu}{^\rho}
+\eta_{\mu\nu}\perpg^{\!\rho\sigma}\nabl_\sigma\alpha\, ,
\hskip 1 cm 
 \Cg_{\mu\nu}{^\rho}\mapsto \Cg_{\mu\nu}{^\rho}\, .\eqn{1.21}\fe}
This  is useful \cite{CSM94} for work like that of 
Vilenkin \cite{Vilenkin91} in a  Robertson-Walker 
cosmological background, which can be obtained from a flat  
spacetime  by a conformal transformation for which 
{\bb {\rm e}^\alpha\fb} is a time dependent Hubble
expansion factor. 


\subsection{Codazzi, Gauss, and Schouten identities}\label{1-5}

As the higher order analogue of (\ref{1.10}) we can go on to introduce the 
{\it third} fundamental tensor$ \cite{C00}$ as
{\be \Xe_{\lambda\mu\nu}{^\rho} \eqdef\etag\,^\sigma_{\ \mu}
\etag\,^\tau_{\ \nu}\perpg^{\!\rho}_{\,\alpha}\onab_\lambda 
\Kg_{\sigma\tau}{^\alpha} \, , 
\eqn{1.22}\fe}
which  by construction is obviously symmetric between the second and third
indices and tangential on all  the first three indices.  In a spacetime 
background that is flat (or of constant curvature as is the case for the 
DeSitter universe model) this third fundamental tensor is fully symmetric 
over all the first three indices by what is interpretable as the {\it 
generalised Codazzi identity}. 

 
In a background with arbitrary Riemann curvature 
{\bb \calR_{\lambda\mu}{^\rho}{_\sigma}\fb}
the {\it generalised Codazzi identity}  is expressible \cite{C00} as 
{\be \Xe_{\lambda\mu\nu}{^\rho}= \Xe_{(\lambda\mu\nu)}{^\rho} +{_2\over^3}
\etag\,^\sigma_{\ \lambda}\etag\,^\tau_{\ {(\mu}}  \etag\,^\alpha_{\ {\nu)}}
\calR_{\sigma\tau}{^\beta}{_\alpha}\perpg^{\!\rho}_{\,\beta}
\  \eqn{1.23}\fe}
A script symbol {\bb \calR\fb} is used here in order to
distinguish the (\nb - dimensional) background Riemann curvature tensor from 
the intrinsic curvature tensor of the (\db - dimensional) 
worldsheet to which the ordinary symbol {\bb \Re\fb}  has already allocated.
For many of the applications that will follow it will be sufficient just to
treat the background spacetime as flat, i.e. to take
{\bb \calR_{\sigma\tau}{^\beta}{_\alpha}=0\, .\fb} 


For $\nb >2$, the background curvature tensor  will be 
decomposible  (if present) in terms of the  background
Ricci tensor and its scalar trace,
{\be \calR_{\mu\nu}= \calR_{\rho\mu}{^\rho}{_\nu}  \, , \hskip 1 cm
 \calR=\calR^\nu_{\ \nu}\, , \eqn{1.24}\fe}
and of its  trace free conformally invariant  Weyl part 
{\bb \calW_{\mu\nu}{^\rho}{_\sigma}\fb}  -- which  can be non zero only for 
$\nb \geq$ 4 -- in the well known \cite{Schouten54} form
{\be\calR_{\mu\nu}{^{\rho\sigma}}=\calW_{\mu\nu}{^{\rho\sigma}} +
{_4\over^{ \nb-2} }
\gb^{[\rho}_{\ [\mu}\calR^{\sigma]}_{\ \nu]}-{_2\over ^{(\nb-1)(\nb-2)} } \calR
\gb^{[\rho}_{\ [\mu}\gb^{\sigma]}_{\ \nu]} \, .\eqn{1.25}\fe}
 In terms of the tangential projection of this background curvature, 
the corresponding {\it internal} curvature tensor  takes the form
{\be  \Re{_{\mu\nu}}{^\rho}{_\sigma}= 2\Kg^\rho{_{[\mu}}{^\tau}
\Kg_{\nu]\sigma\tau}+ \etag\,^\kappa_{\ \mu} \etag\,^\lambda_{\ \nu}
\calR_{\kappa\lambda}{^\alpha}{_\tau} \etag\,^\rho_{\ \alpha}
\etag\,^\tau_{\ \sigma}  \, , \eqn{1.26}\fe}
which is the translation into the present scheme of what is well known in
other schemes as the {\it generalised Gauss identity}. 


The less well known analogue  (attributable  \cite{Schouten54} to 
Schouten) for the (trace free conformally invariant) {\it
outer} curvature   is expressible \cite{C00}
in terms of the relevant projection of the background Weyl tensor as
{\be \Omegag{_{\mu\nu}}{^\rho}{_\sigma}= 2\Cg_{[\mu}{^{\tau\rho}}
\Cg_{\nu]\tau\sigma}
+ \etag\,^\kappa_{\ \mu} \etag\,^\lambda_{\ \nu}
\calW_{\kappa\lambda}{^\alpha}{_\tau}
\perpg^{\!\rho}_{\,\alpha}\perpg^{\!\tau}_{\,\sigma}
 \, . \eqn{1.27}\fe}
In a background that is flat or conformally flat (for which it is 
necessary, and for $\nb \geq 4$ sufficient, 
that the Weyl tensor should vanish) the vanishing of the extrinsic 
conformation tensor {\bb \Cg_{\mu\nu}{^\rho}\fb} will therefore be 
sufficient (independently of the behaviour of the extrinsic curvature 
vector {\bb\Kg^\mu\fb}) for vanishing of the outer curvature tensor 
{\bb \Omegag{_{\mu\nu}}{^\rho}{_\sigma}\, ,\fb} which is the
condition for it to be possible to construct fields of vectors 
{\bb \lambde^\mu\fb}
orthogonal to the surface and such as to satisfy the generalised
Fermi-Walker propagation  condition to the effect that 
{\bb \perpg^{\!\rho}_{\,\mu}
\onab_\nu\lambde_\rho\fb} should vanish.



\section{Laws of motion for a regular brane complex}
\label{Section2}


\subsection{Definition of brane complex}\label{2-1}

The term {\pb}-brane has come  \cite{Achucarroetal87,BarsPope88} 
to mean a dynamic system localised on a timelike support surface of 
dimension {\db=\pb+1\, ,}  in a spacetime background of dimension 
{\nb $>$ \pb}\, . Thus  a zero - brane means  a 
``point particle'', and  a 1-brane means  a 
``string'', while a 2-brane means what is commonly called a ``membrane''. 
At the upper extreme an {\rm ( n-1)}-brane is what 
is commonly referred to as a ``medium'' (as exemplified by a simple fluid).
The codimension-1 (hypersurface supported) case of an {\rm( n-2)}-brane 
(as exemplified by a cosmological domain wall) is what may be referred
to as a ``hypermembrane'', while the codimension-2 case of an 
{\rm ( n-3)}-brane is what may analogously be referred to as a 
``hyperstring''.



A set of branes  forms a ``brane complex'' if the support surface of 
each (\db-1)-brane member is a smoothly imbedded {\db}-dimensional 
timelike submanifold of which the boundary, if any, is a disjoint union 
of support surfaces of  lower dimensional members of the set.  
For the complex to qualify as regular \cite{C00}
it is  required that a 
{\pb}-brane member can act directly only on an {(\pb-1)}-brane 
member on its boundary or on a $(\pb+1)$-brane member on whose boundary 
it is itself located, though it may be passively influenced by
higher dimensional background fields.

Direct mutual interaction between branes with dimension differing by 2
or more would usually lead to divergences, symptomising the breakdown of
a strict -- meaning thin limit -- brane description. To cure that properly, 
a more elaborate treatment -- allowing for finite thickness -- would be 
needed, but it may suffice to use a thin limit approximation \cite{CBU03}
whereby the divergence is absorbed \cite{BCM04,BCM05} in a renormalisation.



\begin{figure}
\centering
\epsfig{figure=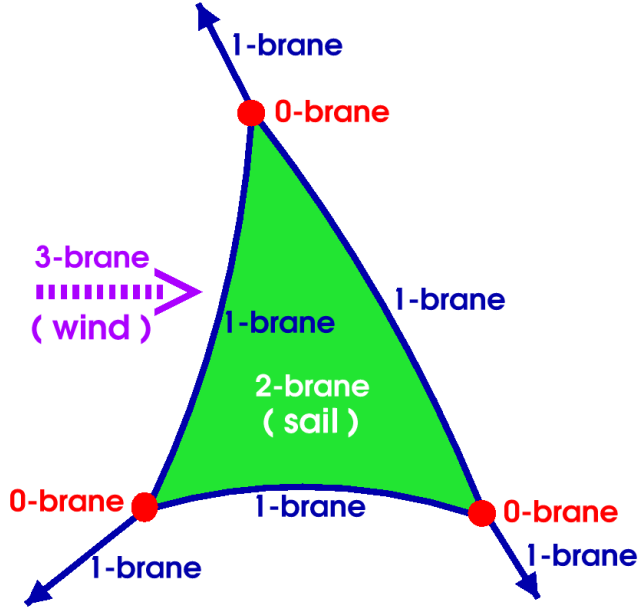, width=10.2 cm}
\caption{
Nautical archetype of a regular brane complex in which a
3-brane (the wind) acts (by pressure discontinuity) 
on  a 2-brane (the sail) hemmed by three 1-branes  (bolt ropes) 
terminating on 0-branes (shackles) that are held in place by  
three more (free) 1-branes (external stay/sheet ropes).}
\label{Fig1}
\end{figure}



In the case of a brane complex, the total action {\bb\Action\fb} will be 
given as a sum of contributions from the various {\rm (d-1)}-branes 
of the complex, of which each has its own Lagrangian {\rm d}-surface 
density scalar {\bb\ud \olag\fb} say. Each supporting {\rm d}-surface 
will be specified by a mapping {\bb \sigme\mapsto x\{\sigme\}\fb} 
giving the local background coordinates {\bb x^\mu\fb} ({\bb\mu\fb}
{\rm \! =\, 0, .... , n-1}) 
as functions of local internal coordinates {\bb \sigme^\ii\ \fb} 
( {\bb \ii\fb} {\rm \!=\, 0, ... , d-1}). The corresponding 
{\rm d}-dimensional surface metric tensor {\bb \ud \hg_{\ii\ji}\fb} 
induced  as the pull back of the {\rm n}-dimensional background 
spacetime metric {\bb \gb_{\mu\nu}\, ,\fb} determines the surface 
measure, {\bb\ud \dr\!\calSg\, ,\fb} in terms of which the total 
action will be expressible as
{\be \Action= \sum_{\db} \int\!\! \ud \dr\!\calSg\, \ud \olag \ ,
 \hskip 1 cm \ud \dr\!\calSg=\sqrt{\Vert^{_{(\db)}} \hg \Vert}\,
 \dr^{\db}\!\sigme \, . \eqn{2.1}\fe}



\subsection{Conserved current and the stress-energy tensor}\label{2-2}

 As well as on its own internal {\rm (d-1)}-brane surface fields and their 
derivatives, and those of any attached {\rm d}-brane, each contribution 
{\bb\ud\olag\fb} will also depend (passively) on 
the spacetime metric $\gb_{\!\mu\nu}$ and perhaps
other background fields, of which the most common example
is a Maxwellian gauge potential 
{\bb \Av_{\!\nu}\, ,\fb}
for which the corresponding field
{\bb \Fr_{\!\mu\nu}=2\nabl_{[\mu}\Av_{\!\nu]}\, ,
\fb} 
 is invariant under gauge changes 
{\bb \Av_{\!\nu}\mapsto\! \Av_{\!\nu}\!+\!\nabl_{\!\nu}\alpha ,\ \fb}  
and is automatically closed,
 {\bb \nabl_{[\rho}\Fr_{\!\mu\nu]}= 0\, , \fb}
 Subject to  the internal dynamic equations of motion given by the 
variational principle stipulating preservation of the action by
variations of the  independent field variables, the effect of 
arbitrary infinitesimal ``Lagrangian'' variations 
{\bb \dL\! \Av_{\!\nu}\, ,\fb}
{\bb \dL\! \gb_{\!\mu\nu}\, ,\fb}  of the background fields will be to 
induce a corresponding variation 
{\be \d \Action\!=\!\sum_{\db}\int\!\! \ud \dr\!\calSg\big(\!
\ud\!\oj{^\mu} \dL\! \Av_{\!\mu}\! 
+\! {_1\over^2} \ud\oT{^{\mu\nu}} 
\dL\!  \gb_{\!\mu\nu}\! \big) \, ,\fe}
from which, for each $(\db\!-\!1)$-brane, one can read out the 
{ electromagnetic surface current} vector {\bb \ud\oj{^{\mu}}\, ,\fb} 
and also (since {\bb \dL (\!\ud \dr\!\calSg)={1\over 2}\ud\etag^{\mu\nu}
(\dL\! \gb_{\!\mu\nu}) \ud \dr\!\calSg\, ,\eqn{varmes}\fb})
the {\it  surface stress momentum energy} tensor  
{\be \ud\oT{^{\mu\nu}}=\ud\oT{^{\nu\mu}}=
2{\partial\!\ud \olag\over\partial\! 
\gb_{\!\mu\nu}}+\ud\olag\ud\eta^{\mu\nu}  
\, . \eqn{2.7}\fe}






For any {\db}-dimensional support surface {\bb \ud\calSg\, ,\fb}  Green's 
theorem gives {\be \int\!\! \ud \dr\!\calSg\ud\onab_\nu\ud\oj{^\nu}=
\oint\! \udi \dr\!\calSg\ud\lambdag_\nu\ud\oj{^\nu}\, ,\eqn{2.11}\fe}
taking the integral on the right  over the boundary {\rm (d-1)}-surface of
{\bb \partial\ud\calSg\fb} of {\bb \ud\calSg\, ,\fb} where 
{\bb\ud\lambdag_\nu\fb} is 
the (uniquely defined) outward directed unit tangent vector on the
{\rm d}-surface at its  {(\db-1)}-dimensional boundary.



The Maxwell gauge invariance condition (independence of {\bb\alpha\fb})
is thus seen to be equivalent to the electric current conservation condition
{\be \up\onab_\mu\up\!\oj{^\mu}=\sum_{\db=\pb+1}\!\!\ud\lambdag_\mu
\ud\!\oj{^\mu}\, , \eqn{2.13}\fe}
which means that the source of charge injection into any particular 
{\rm (p-1)}-brane is the sum of the currents flowing in from the 
{\rm p}-branes to which it is attached. 



\subsection{Force and the stress balance equation}\label{2-4}

The condition of being ``Lagrangian'' means that {\bb \dL \fb} is 
comoving as needed to be meaningful for fields with support  
confined to a particular brane. However for background fields one can also 
define  an ``Eulerian'' variation, {\bb \dE \,  \fb}, with respect to some 
appropriately fixed reference system, in which the infinitesimal
displacement of the brane complex is specified by a vector field 
{\bb \xig^\mu\, .\fb} The difference will be given by
{\bb \dL -\dE=\vec{\ \xig\Libra}\, ,
\eqn{2.19}
\fb}
where the {\bb \vec{\ \xig\Libra}\fb} is the
 Lie differentiation operator, which will be given for the
 relevant background fields by the familiar formulae
{\bb \vec{\ \xi\Libra} \Av_{\!\mu}=\xig^\rho\nabl_\rho  \Av_\mu
\!+\!\Av_{\!\rho}\nabl_\mu\xig^\rho \, , \fb} 
and {\bb \vec{\ \xig\Libra}\! \gb_{\!\mu\nu}\! = \!
2\nabl_{(\mu}\xig_{\nu)}
\, .\fb}



In a fixed Eulerian background, the background fields will have  Lagrangian 
variations  given just by their Lie derivatives with respect to the 
displacement {\bb \xig^\mu\fb}.  Subject to the  internal field equations, 
the action variation {\bb \d \Action\fb} due to the displacement of the 
branes  will therefor just be
{\bb \sum_{\db}\int\!\! \ud \dr\!\calSg\left(\!
\ud\!\oj{^\nu} \vec{\ \xi\Libra} \Av_{\!\nu}\! 
+ {_1\over^2} \ud\oT{^{\mu\nu}}
\vec{\ \xig\Libra}\gb_{\!\mu\nu}\! \right) .\fb}
The postulate that this vanishes for any  {\bb \xig^\mu\fb} entails the
 further {\db}-surface tangentiality restriction 
{\bb \ud\!\!\perpg^\mu_{\ \nu}\ud\oT{^{\nu\rho}}=0\fb} and 
(by the Green theorem) the  dynamic
equations 
{\be \up\onab_\mu\up\oT{^\mu}_\rho= \up \tf_\rho \, , \eqn{2.24}\fe}
in which total force density, 
{\be \up \tf_\rho=\up\of_\rho+\up\cf_\rho \, , \fe}
includes the Faraday-Lorenz  contribution
{\bb \up\of_\rho=\Fr_{\rho\mu}\up\oj{^\mu}
\, ,\fb} from the background, while
on each {(\pb-1)}-brane, the contact force exerted by attached 
{\pb}-branes is
{\be \up\cf_\rho= \sum_{\db=\pb+1}\!\!
\ud\lambdag_\mu\ud\oT{^\mu}_\rho  \, ,\eqn{2.25}\fe}
in which it is to be recalled that, on the {(\pb\!+\!1)}-dimensional
support surface of each attached $\pb$-brane, {\bb \ud\lambdag_\mu\fb}
is the unit vector that is directed normally towards the bounding 
{(\pb-1)}-brane.

The tangential force balance equations will hold as identities when the 
internal field equations are satisfied (because a surface tangential 
displacement has no effect). The non-redundent information governing
the extrinsic motion of a $(\db -\!1)$-brane will be given just by the 
orthogonal  part.  Integrating by parts, as the surface gradient of 
the rank-$(\nb-\db)$ orthogonal projector
{\bb \up\!\!\perpg^{\!\mu}_{\,\nu}\fb} will be given in terms of the 
second fundamental tensor {\bb \up \Kg_{\!\mu\nu}^{\ \ \,\rho}\fb} of 
the $\db$-surface by 
{\be \up\onab_\mu\up\!\!\perpg^{\!\nu}_{\,\rho}=-\up 
\Kg_{\!\mu\nu}^{\ \ \,\rho} -\up \Kg_{\!\mu\ \nu}^{\ \rho} \, ,\fe}
 the extrinsic equations of motion  are finally obtained in the form
{\be \up\oT{^{\mu\nu}}\up \Kg_{\!\mu\nu}^{\ \ \,\rho}=
\up\!\!\perpg^{\!\rho}_{\,\mu}\up \tf^\mu\, .\eqn{2.35}\fe}
 It is to be remarked that this is valid not just for 
a conservative force such as the electromagnetic example considered
above, but also for dissipative forces such as frictional drag\cite{CSM94}
by a relatively moving background medium. 

The most familiar application is to the case $p=1$ of a point particle 
of mass {\bb\mag\fb} with unit velocity vector {\bb\dot x^\mu\fb} 
and orthogonally directed acceleration vector {\bb\ddot x^\mu\, ,\fb}
for which one has {\bb \eta^{\mu\nu}=-\dot x^\mu \dot x^\nu\, ,\fb}
{\bb\oT{^{\mu\nu}}=\mag\,\dot x^\mu \dot x^\nu\, ,\fb}
{\bb\Kg^{\mu\nu\rho}=\dot x^\mu \dot x^\nu\ddot x^\rho\, ,\fb}
so that {\bb\Kg^\rho=-\ddot x^\rho\fb} and 
{\bb \oT{^{\mu\nu}}\Kg_{\!\mu\nu}^{\ \ \,\rho}=\mag\, \ddot x^\rho\, .\fb}


 \section{Canonical Liouville and symplectic currents}
\label{Canonic}


\subsection{Canonical formalism for Branes}\label{3-1}

For the study of small perturbations, and particularly for the 
systematic derivation of conservation laws associated with symmetries,
it is useful to employ a treatment of the canonical kind that was
originally developped in the context of field theory
(as a step towards quantisation) by Witten, Zuckerman, and 
others~\cite{Wi86,CrWi87,Zu87,So94,CF98,Nu00,Ro02}. This section
describes the generalisation of this procedure to 
brane mechanics in the manner initiated by 
Cartas-Fuentevilla~\cite{CF02,CF02b} and developed in collaboration 
with Dani Steer~\cite{CaSte04}. After a general presentation, including 
a review of the relationships between the various (Lagrangian, Eulerian 
and other) relevant kinds of variation, the procedure is illustrated 
by application to a particular category that includes the case of branes of 
purely elastic type.

 
Consider a generic conservative \pb-brane model whose mechanical 
evolution is governed by an action integral of the form
{\be {\Action}=\int {\calL}\, {\rm d}^{\pb+1}\sigme\, ,\label{1e}\fe}
over a supporting worldsheet with internal co-ordinates  
{\bb\sigme^\ii\fb}  {\bb(\ii=0, 1, ... \,\pb )\, ,\fb} and induced 
metric {\bb {\etag}_{\ii\ji}=g_{\mu\nu}x^\mu_{,\ii}x^\nu_{,\ji}\fb}
in a background with coordinates
{\bb x^\mu\, ,\fb \bb(\mu=0, 1, ...\, \nb-1)\, ,\fb} 
{\bb (\nb\geq \pb+1)\ \fb}
and (flat or curved) space-time metric {\bb g_{\mu\nu}\, .\fb} 
The relevant Lagrangian scalar density 
{\bb {\calL}=\Vert{\etag}\Vert^{1/2}{\olag}\, ,
\label{2e}
\fb}
is given as a function of a set of field  components
{\bb \qv^\rmA\ \fb}
 -- including background coords -- and of their surface deriatives, 
{\bb \qv^\rmA_{\, ,\ii} =\partial_\ii\qv^\rmA
=\partial\qv^\rmA/\partial\sigme^\ii\, .\fb}
The relevant field variables  {\bb\qv^\rmA\ \fb} can be of internal or external
kind, the most obvious example of the latter kind being the background
coordinates {\bb x^\mu\fb} themselves.

 The generic action variation,
  {\be \ \ \delta{\calL}= {\calL}_{\!\rmA}\delta\qv^\rmA +
\pred_{\!\rmA}^{\, \ii}\delta\qv^\rmA_{\, ,\ii}\, ,\fe}
specifies partial derivative components {\bb{\calL}_{\!\rmA} \fb}
and and corresponding generalised momentum components
 {\bb{\pred}_{\!\rmA}^{\, \ii}\, .\fb}  The variation principle characterises
dynamically admissible ``on shell'' configurations 
by the vanishing of the Eulerian derivative 
{\be \frac{\delta \calL}{\delta
\qv^\rmA}=\calL_{\!\rmA}-{\pred}_{\!\rmA\, ,\ii}^{\, \ii}\, .\fe}
In terms of this Eulerian derivative, the generic Lagrangian variation
will have the form
{\be \delta\calL=\frac{\delta \calL}{\delta\qv^\rmA}
\delta\qv^\rmA+\big({\pred}_{\!\rmA}^{\, \ii}\delta\qv^\rmA
\big)_{,\ii}\, .\fe}
There will be a corresponding pseudo-Hamiltonian scalar density
{\be\calH= {\pred}_{\!\rmA}^{\, \ii}\qv^\rmA_{\, ,\ii}
-\calL\, ,\fe}
for which
{\be \delta\calH=\qv^\rmA _{\, ,\ii}\delta{\pred}_{\!\rmA}^{\, \ii}
-\calL_\rmA\delta\qv^\rmA      \, .\fe}
(The covariance of such a pseudo-Hamiltonian distingushes it 
from the ordinary kind of Hamiltonian, which depends on the introduction
of some preferred time foliation.)


For an on-shell configuration, i.e. when the dynamical equations
 {\be \frac{\delta \calL}{\delta\qv^\rmA}=0\, ,\fe}
are satisfied, the Lagrangian variation will reduce to a pure surface
divergence,
{\be \delta\calL=\big({\pred}_{\!\rmA}^{\, \ii}\delta\qv^\rmA
\big)_{,\ii}\, ,\fe}
and the correponding on-shell pseudo-Hamiltonian variation will take 
the form
{\be \delta\calH=\qv^\rmA _{\, ,\ii}\delta{\pred}_{\!\rmA}^{\, \ii}
-\pred_{\!\rmA\, ,\ii}^{\, \ii}\delta\qv^\rmA      \, .\fe}


\subsection{Symplectic structure }\label{3-2}

The generic first order variation of the Lagrangian will be given by
{\be \delta\calL=\frac{\delta \calL}{\delta\qv^\rmA}
\delta\qv^\rmA+\varthetar^\ii_{\, ,\ii}\, .\fe}
in terms of the generalised Liouville 1-form
(on the configuration space cotangent bundle) defined  by
{\be \ \ \varthetar^\ii=\pred_\rmA^{\, \ii}\delta \qv^\rmA\, .\fe}

Now consider  a pair of successive independent variations 
{\bb \deI\, ,\fb \bb \deJ\, , \fb} which will give a 
second order variation of the form
{\be \deJ\deI\calL=\deJ\Big(\frac{\delta \calL}{\delta\qv^\rmA}\Big)
\deI\qv^\rmA\!+\!\frac{\delta \calL}{\delta\qv^\rmA}\deJ\deI \qv^\rmA
\!+\!\big(\deJ\pred_\rmA^{\, \ii}\deI\qv^\rmA\!+\!
\pred_\rmA^{\, \ii}\deJ\deI\qv^\rmA\big)_{,\ii}\, .\fe}
Thus using the commutation relation {\bb \deJ\deI=\deI\deJ\fb} one gets
{\be
\deJ\Big(\frac{\delta \calL}{\delta\qv^\rmA}\Big)\deI\qv^\rmA-
\deI\Big(\frac{\delta \calL}{\delta\qv^\rmA}\Big)\deJ\qv^\rmA
=\varpirIJ^\ii_{\, ,\ii}\, ,\fe}
where the symplectic 2-form (on the configuration space cotangent bundle)
is defined by
{\be \ \ \varpirIJ^\ii= \deI\pred_\rmA^{\, \ii}\deJ\qv^\rmA-
\deJ\pred_\rmA^{\, \ii}\deI\qv^\rmA\, .\label{ome}\fe}

For an on-shell perturbation we thus obtain
{\be\frac{\delta \calL}{\delta\qv^\rmA}=0\hskip 1 cm \Rightarrow\
\hskip 1 cm \delta\calL=\varthetar^\ii_{\, ,\ii}\,,\fe}
while for a pair of on-shell perturbations we obtain
{\be\deI\Big(\frac{\delta \calL}{\delta\qv^\rmA}\Big)=
\deJ\Big(\frac{\delta \calL}{\delta\qv^\rmA}\Big)=0 \hskip 1cm
\ \Rightarrow\hskip 1cm \varpirIJ^\ii_{\, ,\ii}=0\, .\fe}



The foregoing surface current conservation law is
expressible in shorthand as
{\be\varpir^\ii_{\, ,\ii}=0 \, ,\fe}
 in which the closed (since manifestly exact) symplectic 2-form
(\ref{ome}) is specified in concise wedge product notation as
{\be \varpir^\ii=\delta\wedge\varthetar^\ii=\delta
\pred_\rmA^{\, \ii}\wedge\delta \qv^\rmA\, .\fe}
Some authors prefer to use an even more
concise notation system in which it is not just the relevant distinguishing
(in our case acute and grave accent) indices that are omitted but
even the wedge symbol {\bb\wedge\fb} that indicates the antisymmetrised
product relation. However such an extreme  level of abbreviation is 
dangerous~\cite{CF02} in contexts in which symmetric products
are also involved. 



\subsection{Translation into strictly tensorial form}\label{3-3}

To avoid the gauge dependence involved in the use of
auxiliary structures such as local frames and internal surface
coordinates, by working ~\cite{Ca93} just with quantities that 
are strictly tensorial with respect to the background space, one needs to
replace  the surface current densities whose components {\bb
\varthetar^\ii\fb} and {\bb \varpir^\ii\fb} depend on the choice of
the internal coordinates  {\bb\sigme^\ii ,\fb} by vectorial quantities with 
 strictly tensorial background coordinate components given by
{\be  \Thetar^\nu= \Vert{\etag}\Vert^{- 1/2} x^\nu_{\, ,\ii}
\varthetar^\ii \, ,\hskip 1 cm
\Omegar^\nu=\Vert{\etag}\Vert^{- 1/2} x^\nu_{\, ,\ii}\varpir^\ii\, .\fe}
and with strictly scalar divergences given by
{\be\overline\nablag_{\!\nu}\Thetar^\nu=\Vert\etag\Vert^{-1/2}
\varthetar^\ii_{\, ,\ii}\, ,\hskip 1 cm
\overline\nablag_{\!\nu}\Omegar^\nu=\Vert\etag\Vert^{-1/2}
\varpir^\ii_{\, ,\ii}\, .\fe}


In terms of  the surface projected covariant 
differentiation operator defined in terms of the fundamental tensor 
{\bb \etag^{\mu\nu}=\etag^{\ii\ji} x^\mu_{\, ,\ii}x^\nu_{\, ,\ji}\fb}
by {\bb\ \ \overline\nablag_{\!\nu}=\etag^\mu_{\ \nu}\nablab_{\!\mu}\, ,\fb} 
one thus obtains a Liouville current conservation law of the form
{\be\overline\nablag_{\!\nu}\Thetar^\nu=0\fe}
for any symmetry generating perturbation, 
i.e. for any infinitesimal variation {\bb\delta\qv^\rmA\fb} 
such that {\bb\delta\calL=0\, .\fb}
Similarly a symplectic current conservation law of the form
{\be\overline\nablag_{\!\nu}\Omegar^\nu=0\fe}
 will hold for any pair of perturbations that are on-shell, i.e.
such that {\bb\delta(\delta\calL/\delta \qv^\rmA)=0\, .\fb}



\subsection{Application to hyperelastic case}\label{3-6}

In typical applications, the relevant set of configuration  components 
{\bb \qv^\rmA\fb} will include a set of brane field components 
{\bb \varphiv^\alphab\fb} as well as the background coords {\bb x^\mu\fb}, 
so that in terms of displacement vector {\bb\xig^\mu=\delta x^\mu\fb} 
the Liouville current will take the form
{\be \Thetar^\nu=\Vert\etag\Vert^{-1/2} x^\nu_{\, ,\ii}\big(
\pred_\alphab^{\ii}\,\delta\varphiv^\alphab+
\pred_\mu^{\ \ii}\,\xig^\mu\big)=
\pired_\alphab^{\nu}\,\delta \varphiv^\alphab+
\pired_\mu^{\ \nu}\,\xig^\mu\, ,\fe}
in which the latter version replaces the original momentum components
by the corresponding background tensorial momentum variables, which are 
given by
{\bb \pired_\alphab^{\,\nu}=\Vert\etag\Vert^{-1/2}\, x^\nu_{\, ,\ii}\,
\pred_\alphab^{\ii}\fb}
and
{\bb \pired_{\!\mu}^{\ \nu}=\Vert\etag\Vert^{-1/2}\, x^\nu_{\, ,\ii}\,
\pred_\mu^{\ \ii }\, .\fb}



The \underline{hyperelastic} category \cite{Ca07} (generalising the 
case of an ordinary elastic solid which includes the special case of 
an ordinary barotropic perfect fluid) consists of brane models 
in which -- with respect to a suitably
comoving internal reference system {\bb \sigme^\ii\fb} -- there are no
independent surface fields at all --  meaning that the
{\bb \varphiv^\alphab\fb} and the {\bb\pred_\alphab^{\,\ii}\fb} are 
absent -- and in which the only relevant background field is the 
metric {\bb \gb_{\mu\nu}\fb} that is specified as a function of the 
external coordinates {\bb x^\mu\fb}. In any such  case, 
the generic variation of the Lagrangian is determined just by 
the surface stress momentum energy density tensor
{\bb\overline\Tr{^{\mu\nu}}\fb} according to the standard prescription 
{\be \delta\calL=\frac{1}{2}\Vert\etag\Vert^{1/2}\,
\overline\Tr^{\mu\nu}\,\dL \gb_{\mu\nu}\, ,\fe} 
whereby {\bb\overline\Tr{^{\mu\nu}}\fb} is specified in terms of partial
derivation of the action density with respect to the metric.  


In a fixed background (i.e. in the absence of any Eulerian variation 
of the metric) the Lagrangian variation of the metric will be given by
{\bb \dL \gb_{\mu\nu}=\vec\xig\Libra \gb_{\mu\nu}
=2\nablab_{\!(\mu}\xig_{\nu)}\, .\fb}
Comparing this to  canonical prescription
{\bb\delta\calL=\calL_\mu\xig^\mu
+\pred_\mu^{\ \ii}\xig^\mu_{\, ,\ii}\fb}
with {\bb\xig^\mu=\delta x^\mu\fb}
shows that the relevant partial derivatives will be given by
the (non-tensorial) formulae
{\bb\calL_\mu=\Vert\etag\Vert^{1/2}\,\Gammab_{\!\mu\ \rho}^{\ \nu}
\overline\Tr_{\!\nu}{^\rho}\, ,\fb} {\bb
\pred_\mu^{\ \ii}=
\Vert\etag\Vert^{1/2}\,\overline \Tr_{\mu\nu}
\etag^{\ii\ji} x^\nu_{\, ,\ji} \, .\fb}
It can thus be seen  that in the \underline{hyperelastic}
case, the canonical momentum tensor {\bb \pired_\mu{^{\ \nu}}\fb}
and the Liouville current {\bb\Thetar^\nu\fb}
will be given just in terms of surface stress tensor
{\bb\overline{\Tr}{^{\mu\nu}}\fb} by the very simple formulae
{\be \pired_\mu{^{\nu}}=\overline{\Tr}_\mu{^{\nu}}\, ,\hskip 1 cm
\Thetar^\nu=\overline{\Tr}_\mu{^{\nu}}\xig^\mu\, .\fe}


In order to proceed, we must consider the second order metric
variation, whereby (following Friedman and Schutz~\cite{FrSc75})
the hyper Cauchy tensor (generalised elasticity tensor)
{\bb \overline{\Cauch}{^{\mu\nu\rho\sigma}}
=\overline{\Cauch}{^{\rho\sigma\mu\nu}}\fb}
is specified~\cite{BC95} in terms of Lagrangian variations by
a partial derivative relation of the form
{\be\dL\big(\Vert\etag\Vert^{1/2}\, \overline{\Tr}{^{\mu\nu}}\big)=
\Vert\etag\Vert^{1/2}\overline{\Cauch}{^{\mu\nu\rho\sigma}}
\dL \gb_{\rho\sigma}\, .\label{hyperC}\fe}
The symplectic current is thereby obtained in the form
{\be \Omegar^\nu= \OP_\mu^{\ \nu}
\wedge \xig^\mu\, , \label{elas} \fe}
where
{\be \OP_\mu^{\ \nu}=2\overline{\Cauch}{_{\mu\ \rho}^{\ \, \nu\ \sigma}}
\overline\nablag_{\!\sigma}\xig^\rho+\overline{\Tr}{^{\nu\rho}}
\overline\nablag_{\!\rho}\xig_\mu\, .\fe}








 \section{Brane perturbation by  gravitational radiation}
\label{extRad}

\subsection{Generic case}\label{4-75}

A background metric perturbation
{\bb \delta \gb_{\mu\nu}= \hb_{\mu\nu}\fb}
will provide an extra Lagrangian and stress contributions
{\bb\delta\olag=
{_1\over^2}\overline\Tr{^{\rho\sigma}}\hb_{\rho\sigma}\, ,\fb} and
{\bb \delta\overline\Tr{^{\mu\nu}}=\overline{\Cauch}{^{\mu\nu\rho\sigma}}
\hb_{\rho\sigma}\, ,\fb}
whence  a corresponding force increment
{\bb \delta \overline \tf{^\mu}={_1\over^2}\overline \Tr{^{\nu\sigma}}
\nab^\mu \hb_{\nu\sigma} - \overline\nablag_{\!\nu}\big( \overline
\Tr{^{\nu\sigma}} \hb_\sigma{^\mu} \big) \, . \fb}
The effect of this is expressible as the inclusion of an extra
term  {\bb\overline \tf_{_{\!\rm G}}{^\mu}\fb} on the right of the
original force balance equation, as expressed in terms of the
unperturbed values of the metric  {\bb\gb_{\mu\nu}\, ,\fb} stress tensor 
{\bb\overline\Tr{^{\mu\nu}}\, ,\fb} and force density 
{\bb \tf{^\mu}\, ,\fb} so as to obtain a perturbed force balance of the 
form
{\be \onab_\nu\Tr{^{\nu\mu}}= \tf^\mu+\overline \tf_{_{\!\rm G}}{^\mu}
\, , \label{pertforce}\fe}
in which the effective gravitational perturbation contribution is given by 
{\be \overline \tf_{_{\!\rm G}}{^\mu}={_1\over^2}\overline \Tr{^{\nu\sigma}}
\nab^\mu \hb_{\nu\sigma} - \overline\nablag_{\!\nu}\big( \overline
\Tr{^{\nu\sigma}} \hb_\sigma{^\mu}+ \overline{\Cauch}{^{\mu\nu\rho\sigma}}
\hb_{\rho\sigma} \big) \, , \eqn{22a}\fe}
a formula that was not so well known until relatively recently~\cite{BC95}.


\subsection{The case of a simple Dirac-Nambu-Goto type brane}
\label{4-7}
 
The simplest dimensionally unrestricted application,  is to a \pb-brane
of the Dirac-Nambu-Goto type, for which the relevant master
function is simply constant, so given in terms of a corresponding
Kibble mass  {\bb \hbox{\smM}_{_{\rm K}}\fb} by
{\bb \olag= -\hbox{\smM}_{_{\rm K}}^{\,\pb+1} \, .\eqn{70}\fb}
(In the context of superstring theory {\bb \hbox{\smM}_{_{\rm K}}\fb}
is  typically of the order of magnitude of the Planck mass
$\hbox{\smM}_{_{\rm P}}$, whereas in the context of cosmic string theory
the Kibble mass is expected to comparable with the relevant
Higgs mass, {\bb \hbox{\smM}_{_{\rm X}}\, .\fb})
In this special case, the surface stress momentum energy tensor is of 
course simply proportional to the fundamental tensor:
{\be \overline \Tr{^{\mu\nu}}=-\,\calT\,
\etag^{\mu\nu}  \, ,\hskip 1 cm  \calT=
\hbox{\smM}_{_{\rm K}}^{\,\pb+1}
\eqn{71}\fe}
so its trace will be given by {\bb \ \overline \Tr
=-\fb}{\color{bk}(\pb+1)}{\bb \,\calT
\, ,\fb} where {\bb\calT\fb} is interpretable as the surface tension.
The corresponding the hyper-Cauchy tensor is found\cite{BC95} to be
{\be \overline{\Cauch}{^{\mu\nu\rho\sigma}}=
\calT
\big(\etag^{\mu(\rho}\etag^{\sigma)\nu}-\textstyle{1\over 2}
\etag^{\mu\nu}\etag^{\rho\sigma}\big)\, .\eqn{71A}\fe}
The dynamical equation of motion (\ref{pertforce})
will therefor
 reduce to the form
{\be \calT \,\Kg^\rho = - \tf{^\rho}\, ,
\eqn{75}
\fe}
in which (as well as the possibility of drag) the right hand side 
will include an effective  gravitational contribution  
expressible\cite{BC95} in the form
{\bb \tf_{_{\!\rm G}}{^\mu}= \tf_{_{\!\rm I}}{^\mu}+
\tf_{_{\!\rm I\! I}}{^\mu}\, ,
\fb}
with
{\be \tf_{_{\!\rm I}}{^\mu}=\calT\,
\perpg^{\mu\nu}\etag^{\rho\sigma}\big{(} \nab_{\rho}\hb_{\nu\sigma}-
\textstyle{1\over 2}\nab_{\nu}\hb_{\rho\sigma}\big{)}\, ,\eqn{77}\fe}
{\be \tf_{_{\!\rm I\! I}}{^\mu}= \calT\,
\big{(}\perpg^{\mu\nu}\Kg^{\rho}+\textstyle{1\over 2}\etag^{\rho\nu}
\Kg^{\mu}-\Kg^{\nu\rho\mu}\big{)}\hb_{\nu\rho}\, .\eqn{78}\fe}
It was observed by Battye\cite{B95,CB98} that the early work on
gravitational perturbations of strings cited by  Vilenkin and 
Shellard in their 1994 treatise~\cite{VS94} was seriously flawed 
by the use for estimating {\bb \tf_{_{\!\rm G}}{^\mu}\fb} of a 
formula (7.7.3) without the orthogonal projection operator 
{\bb \perpg^{\mu\nu}\fb} in the expression (\ref{77}) for  
{\bb \tf_{_{\!\rm I}}{^\mu}\, ,\fb} and entirely  lacking the 
contribution{\bb \tf_{_{\!\rm I\! I}}{^\mu}\fb} which might be 
relatively negligible for high frequency radiation\cite{B95} of 
external origin, but not in the case of self-interaction for 
which the two contributions will be comparable. The self
interaction contributions from (\ref{77}) and (\ref{78}) will
be separately divergent, but in the ``hyperstring'' case these 
divergences will actually cancel each other. Thus (contrary to 
what was claimed in (7.7.7)~\cite{VS94}) the total
self-interaction will remain finite\cite{CB98,BCM04,BCM05} 
whenever the codimension is 2, as for an ordinary  string 
in 4 dimensions (or for a ``brane-world'' in 6 dimensions).
\vfill\eject

\subsection{Regularisation of  self-interaction}

To treat such self-interaction one must face the problem that the 
regularity condition (see Figure \ref{Fig1}) is violated whenever 
a brane of dimension {\bb \db=\pb+1\fb} acts on a background of 
dimension {\bb \nb \geq\db+2\, ,\fb}. To cure this, a physically 
realistic regularisation involves replacing the infinitely 
thin worldsheet by a support of finite thickness. The divergent 
self-interaction fields such as {\bb\Av_{\!\mu}\fb} and 
{\bb \hb_{\mu\nu}\fb} are then replaced by regularised averages 
{\bb\widehat\Av_{\!\mu}\fb} and {\bb \widehat \hb_{\mu\nu}\fb} with 
dominant contribution proportional to the relevant source 
\cite{BCM04,BCM05}. This means {\bb \widehat\Av_{\!\mu}\propto 
\oj{^{\mu}}\fb} and {\bb \widehat \hb_{\mu\nu}\propto 
(\nb-2)\oT{_{\!\mu\nu}}-\oT{_{\!\sigma}^{\ \sigma}}\gb_{\mu\nu} 
\, , \fb} which for a Nambu-Goto hyperstring, 
{\bb \pb=\nb-3\, ,\fb} gives {\bb \widehat \hb{^{\mu\nu}}
\propto (\pb+1)\calT\perpg^{\!\mu\nu}\, ,\fb}  with a 
proportionality coefficient that diverges as the thickness tends to 
zero. On such world sheet confined fields, the ordinary gradient 
operator {\bb\nabl_\nu \fb} must  be replaced by the corresponding 
regularised operator {\bb\widehat \nablag_{\!\nu}\, ,\fb} so that for 
example the field {\bb \Fr_{\!\mu\nu}=2\nabl_{[\mu}\Av_{\!\nu]}\fb} 
will have the regularised average {\bb \widehat \Fr_{\!\mu\nu}= 
2\widehat\nablag_{[\mu}\widehat\Av_{\!\nu]}\, ,\fb} as needed
for the electromagnetic self-interaction force density 
 {\bb \widehat\tf_\rho=\widehat\Fr_{\rho\mu}\oj{^\mu}\, . \fb}
The required result, giving zero gravitational contribution, 
{\bb \widehat\tf_{_{\!\rm G}}{^\mu}=0\, , \fb} for
Nambu-Goto hyperstrings (including \cite{CB98} the ordinary string case \pb=1
with \nb=4) has been shown \cite{CBU03} to be
provided generally by the conveniently simple 
and easily memorable formula
{\bb\widehat \nablag_{\!\nu}=\onab_\nu+\frac{_1}{^2} \Kg_\nu\, .\fb}

\section{Vorton equilibrium states of elastic string 
loops}\label{Vortons}


\subsection{The category of simple elastic string models}\label{6-2}

For any string model the fundamental tensor 
of the 2 dimensional worldsheet will be expressible in terms of
any orthonormal diad of space like and timelike vectors
{\bb \ugn^\nu\, ,\widetilde\ugn{^\nu}\fb} as 
{\bb \etag^\mu_{\ \nu}=-\ugn^\mu\ugn^\nu+\widetilde\ugn
{^\mu}\widetilde\ugn{^\nu}\, . \fb} 
There will generically be a prefered diad 
with respect to which the symmetric surface stress 
energy tensor will be expressible as
{\be \oT{^{\mu\nu}}=\calU \ugn^\mu\ugn^\nu-\calT\,\widetilde\ugn{^\mu}\,
\widetilde \ugn{^\nu} \label{TU}\fe} 
where {\bb \calT\fb} is the 
string tension, and {\bb \calU\fb} is the surface energy density, which,
in the elastic case, is determined as a function of {\bb \calT\fb}
by an equation of state.

In addition to the extrinsic (transversely polarised)``wiggle'' perturbations 
which, as in any string model, travel with a characteristic velocity
 {\bb\vv=\sqrt{\calT/\calU}\fb} such a model has perturbation modes of
only one other kind: these are sound type (longitudinal compression) 
``woggle'' modes, which propagate relative to the locally
preferred frame with speed given by the formula
{\bb\vv_{_{\rm L}}=\sqrt{ -{\rm d}\calT/{\rm d}\calU}\fb}.
A particularly important special case is that of models of the
integrable transonic type \cite{C95} for which 
the ``wiggle'' and ``woggle''
speeds coincide, which occurs when the equation of state is specified
simply by the specification of a fixed value for the product 
{\bb \calU \calT\fb}. The kind of 
model appropriate for representing such familiar technical
applications as bow strings, or the strings of musical instruments,
will generally be of subsonic type, meaning that the wiggle speed  
{\bb\vv\fb} is less than the sonic speed {\bb \vv_{_{\rm L}} \fb},
while on the other hand it has been shown by Peter \cite{P92} 
that models of supersonic type will commonly be needed for the 
representation of cosmic strings of the conducting vacuum vortex 
type envisaged by Witten \cite{Witten85}. 






A model of any such elastic type is specifiable in variational form by a
string Lagrangian {\bb \Lr \fb} depending only on the magnitude of 
the gradient of some stream function {\bb \varphiv\fb} (which in the
Witten case represents the phase of a complex scalar field). This
means that the string model is characterised by a single variable equation 
of state giving {\bb \Lr \fb} as a function of the scalar
 {\bb \vww=\hg^{\ii\ji}\varphiv_{,\ii} \varphiv_{,\ji}\, .\fb}
It is useful \cite{C95a,CarPeGan97} to introduce the corresponding 
adjoint formulation in terms of the quantity 
{\bb \duLag=\Lr+\vww\rkap\, ,\fb}
with {\bb \rkap=-2\,{{\rm d}\Lr}/{{\rm d}\vww}\, .
\fb} 
When {\bb\vww<0\fb}, one finds that 
the tension and energy density will be given by
{\bb\calT=- \Lr\, , \hskip 1 cm \calU=-\duLag\, ,\fb}  
while when {\bb\vww>0 \fb} they will be given by
{\bb\calT=- \duLag\, ,\hskip 1 cm  \calU=-\Lr\, .\fb} 
In all cases the phase gradient is proportional to a surface current,
{\bb\ocr^\mu=x^\mu_{\, ,\ii}\ocr^\ii\, ,\fb} {\bb
\ocr^\ii=\rkap\hg^{\ii\ji}\varphiv_{,\ji}=-{\partial\Lr}/{
\partial\varphiv_{,\ii}}\, ,\fb} that has the property of being conserved,
{\bb (\sqrt{-\hg}\, \ocr^\ii)_{,\ii}=0\, ,\fb}
whenever there is no external force, so that the equation of motion
of the worldsheet reduces to the simple form
{\bb \overline\Tr{^{\mu\nu}}\Kg_{\mu\nu}{^\rho}=0
\, ,\label{freemov}\fb}
with
{\bb \overline\Tr{^{\mu\nu}}=2\rkap^{-1}\ocr^\mu\ocr^\nu
+\Lr\etag^{\mu\nu}\, .\fb}


 When he originally introduced the concept of conducting cosmic strings
\cite{Witten85} Witten suggested that a simple
linear action formula, {\bb \Lr=-\mm^2(1+\delth_\ast^{\,2}\, \vww)
\, ,\fb} involving just a single extra parameter (namely a lengthscale 
{\bb \delth_\ast\fb}) might be used as a good approximation,
 least in the weak current limit for which {\bb 
\vww\fb} is sufficiently small. However it subsequently became
 clear that such a linear formula
is inadequate even in the weak current limit, since it
implies that wiggle propagation would always be subsonic
{\bb \vv^2 < \vv_{_{\rm L}}^{\, 2}\fb}, whereas detailed examination
of the relevant kind of vacuum vortex by Peter \cite{P92} 
revealed that the wiggle propagation in such a case would typically
be supersonic {\bb \vv^2>\vv_{_{\rm L}}^{\, 2}\fb}
As a more satisfactory replacement for Witten's direct linearity 
ansatz, it has been found \cite{CP95,HC08}
that at the cost of introducing one more mass scale
{\bb \mm_\star\, ,\fb} a reasonably good representation is obtainable
by using an ansatz of logarithmic form
{\bb \Lr=-\mm^2\!-\frac{_1}{^2}\mm_\star^{\,2}\, {\rm ln}\,
\{1\!+\!\delth_\star^{\,2}\,\vww\}\, .\fb}





\subsection{Stationary string states in flat background}\label{7-4}

We shall conclude this overview by considering what can be
said about stationary equilibrium states, as characterised,
in a flat background  a world sheet that is tangent to 
a timelike unit static Killing vector satisfying
{\bb \nablab_{\!\mu} k^\nu=0\fb}. In such a worldsheet
there will also be an orthogonal  (and therefor spacelike)
unit tangent vector {\bb \ebl^\mu\fb} satisfying
the invariance condition
{\bb k^\nu \nablab_{\!\nu} \ebl^\mu=0 \, .\fb}
For such a worldsheet, the first fundamental tensor 
will be given by
{\bb \etag^{\mu\nu}=-k^\mu k^\nu
+\ebl^\mu \ebl^\nu\, ,\fb} 
while in terms of the curvature vector,
{\bb \Kg^\mu=
\ebl^\nu\nablab_{\!\nu} \ebl^\mu\, ,\fb}
the second fundamental tensor will be given by
{\bb \Kg_{\mu\nu}{^\rho}= \ebl_\mu \ebl_\nu \Kg^\rho\, .\fb}

Within the worldsheet, the preferred timelike eigenvector
of the stress energy tensor, as characterised by the relation
{\bb \overline\Tr{^\mu_{\, \nu}}\uvec^\nu=-\calU \uvec^\nu ,\fb}
will be expressible in the form
 {\be \uvec^\mu=(1\!-\!\vv^2)^{-1/2}
( k^\mu\!+\!\vv\,\ebl^\mu)\fe} 
which defines the relative flow velocity {\bb \vv\, .\fb}  
Under these conditions,  the free dynamical equation 
(\ref{freemov}) can be seen to
reduce to the simple form
{\bb (\calU-\vv^2\calT)\Kg^\rho=0\, .
\label{eqm}
\fb}

For an infinitely long string this equation can of course
be solved in a trivial manner by choosing a configuration
that is straight, which means {\bb\Kg^\rho=0 \ ,\fb}
in which case the value of {\bb \vv \fb} is unrestricted.
However for a finite closed loop the curvature cannot vanish 
everywhere, and where {\bb\Kg^\rho\fb} is non-zero the only way 
of satisfying the extrinsic equilibrium condition(\ref{eqm}) is 
for the relative flow velocity to bethe same as the relevant 
wiggle propagation speed: {\bb \vv=\sqrt{\calT/\calU}\, ,\fb}
while to satisfy the intrinsic (current conservation) 
equilibrium condition it is trivially sufficient  (and 
generically necesssary) for the value of this speed to be 
uniform. Provided this centrifugal equilibrium condition is 
satisfied, there is no retriction on the curvature, which need 
not be uniform: thus the equilibrium configuration of the string 
loop need not be circular, but may have an arbitrary shape.

After thus obtaining the generic condition for string loop 
equilibrium, the next problem is to find which of such  
vorton equilibrium states are stable. This question has so far 
been dealt with \cite{CarMar93,Mar94} only in the simple case of
equilibrium configurations that are circular.



\subsection{Stability criterion for circular vorton states}\label{7-5}

It is easy to see that the stability of a uniform circular 
equilibrium state of an elastic string loop in a flat background will
depend just on the extrinsic (wiggle type) and longitudinal (sound 
type) perturbation speeds,  {\bb\vv\fb} and {\bb\vv_{_{\rm L}}\fb}. 
Moreover it is fairly easy to show \cite{CarMar93} that such a state 
will always be stable in the subsonic case, 
{\bb \vv^2\leq \vv_{_{\rm L}}^{\,2}\fb},
which is what is most likely to be relevant in a terrestrial 
engineering context.Even in the supersonic case, it has been 
shown \cite{CarMar93} that monopole {\bb \nN=0\fb} and dipole 
{\bb \nN=1\fb} perturbation modes are always stable.  However 
instability may occur for higher modes, {\bb \nN\geq 2 \fb} for 
which, in a state with radius {\bb \aa\, ,\fb} the eigenfrequency 
{\bb \omegam\fb} is given by the solution of an equation of the cubic 
form {\bb \xm^3+\bv_2 \xm^2+\bv_1 \xm+\bv_0=0\, ,\fb} for the 
quantity {\bb \xm={\aa\,\omegam}/{\vv_{\!_+}\,\nN}\, ,\fb} where
{\fb \vv_{\!_+}={2\vv}/(1+\vv^2) \, ,\fb} is the relative 
velocity of orthogonaly polarised forward propagating wiggles,
and the coefficients of the cubic are given by 
{\bb \bv_2=\Ggamma\!-2-\xxxi\, ,\fb}  {\bb\bv_1=
-2\Ggamma\!+(1+\xxxi)\left(1\!-\nN^{-2}\right)\, ,\fb}
 {\bb \bv_0=\Ggamma\left(1\!-\nN^{-2}\right)\, ,\fb}
using the notation
{\bb \xxxi=\Ggamma(1-\vv_{\!_+}^{\,2})\, ,\fb} {\bb
\Ggamma={\vv_{\!+}^{- 2}{( \vv_{_{\rm L}}^{\,2}-\vv^2)}/
(1-\vv_{_{\rm L}}^{\,2}\,\vv^2)}\, .\fb}

The stability criterion, for all the roots to be real, is the 
positivity of a discriminant
{\bb \Ddelta=\bv_2^{\,2}\bv_1^{\,2}+ 18\bv_2\bv_1\bv_0-
4\bv_1^{\,3}-4\bv_2^{\,3}\bv_0-27\bv_0^{\,2}\, .\fb}
Figure \ref{Fig2} shows the zones of negativity (instability) 
for the lowest relevant mode numbers, {\bb \nN=2,3, ...\fb} 
by Martin \cite{Mar94}. In the ultrarelativistic limit 
{\bb \vv\rightarrow 1\fb}, {\bb \vv_{_{\rm L}}\rightarrow 1\fb}  
that is relevant for weak currents in conducting cosmic strings, 
one gets {\bb \xxxi\!\rightarrow 0 \, \fb} and
{\be \Ddelta\!\rightarrow 4\nN^{-2}(\Ggamma\!+\!1\!+\!\nN^{-1})^2
(\Ggamma\!+\!1\!-\!\nN^{-1})^2 ,\fe}
which is strictly positive (implying stability) almost always, the 
unstable exceptions being on the lines converging in the plot to 
the limit point {\bb \vv^2=1 \fb}, {\bb \vv_{_{\rm L}}^{\,2}=1\fb} 
with gradient given in terms of the corresponding node number by 
{\bb 1/(2\nN-1)\, .\fb}

The upshot is that although some circular vorton states are unstable,
there are plenty more -- the ones that would presumably be 
selected under natural conditions -- that are stable, at least
with respect to macroscopic string perturbations. 
It is however to be remarked that -- since it deals only with
the thin string limit -- the kind of analysis described here
can not resolve the (sensitively model dependent) issue of stability with
respect to quantum effects or other processes involving the
microscopic internal structure of the vacuum vortex or whatever
else may constitute the string.

\end{document}